# Phenotypic Performance of *Kambro* Crossbreeds of Female Broiler Cobb 500 and Male *Pelung Blirik Hitam*

(PERFORMA FENOTIPIK KAMBRO HASIL PERSILANGAN ANTARA BETINA BROILER COBB 500 DAN JANTAN PELUNG BLIRIK HITAM)

**I Wayan Swarautama Mahardhika\*, Budi Setiadi Daryono**

Gama Ayam Research Team, Laboratory of Genetics and Breeding, Faculty of Biology, Gadjah Mada University, Jl. Teknika Selatan, Sinduadi, Mlati, Kabupaten Sleman, Daerah Istimewa Yogyakarta, Indonesia 55281.  
\*Email: *i.wayan.sm@mail.ugm.ac.id*

## ABSTRACT

This research was conducted to measure the phenotypic performance of *Kambro* crossbreeds of *Pelung Blirik Hitam* and Broiler Cobb 500. Based on Body Weight (BT) measurement, *Kambro* population (n = 17) has an average BT of 1,244.14 ± 453.82 grams significant (p <0.01) to $F_1$ *Pelung* (n = 7) with an average BT of 602.88 ± 79.93 grams in 8 weeks period with *ad libitum* diet of standard feed. Phenotypic performance of *Kambro* significant to $F_1$ *Pelung* based on the measurement of linear body weight parameter, vitality parameter, PPa-PBe parameter and phenotype parameter. *Kambro* has the phenotype combination of parental generation based on phenotype parameter. PPa parameter was suitable BT estimation model based on non-linear quadratic regression (r = 0.956) with formula 1.84E3 ±3.54E2*x+31.73*$x^2$. Difference between chicken group (p<0.014) was significant to BT and interaction between group and linear body weight parameter was not significant based on Analysis of Covariance. Mortality rate of *Kambro* was lower than $F_1$ *Pelung* with the absent of vaccination in semi-intensive rearing system. As the size of hybrids population was limited, research findings must be validated with larger population size of hybrids.

## ABSTRAK

Riset ini diadakan dengan tujuan mengukur performa fenotipik Kambro hasil persilangan antara Pelung Blirik Hitam dan Broiler Cobb 500. Berdasarkan pengukuran Bobot Tubuh (BT), rerata BT Kambro (n = 17) mencapai 1.244,14 ± 453,82 gram signifikan (p<0,001) terhadap $F_1$ Pelung (n = 7) dengan rerata BT 602,88 ± 79,93 gram pada umur 8 minggu dengan diet pakan standar ad libitum. Performa fenotipik Kambro signifikan terhadap $F_1$ Pelung berdasarkan parameter bobot tubuh linear, parameter vitalitas, parameter PPa-PBe dan parameter fenotipe. Kambro memiliki perpaduan fenotip indukannya berdasarkan parameter fenotipe. Parameter PPa merupakan model estimasi BT Kambro berdasarkan regresi *non-linear quadratic* (r = 0,956) dengan formula 1.84E3 ± 3.54E2*x+31.73*$x^2$. Perbedaan grup antar grup signifikan (p<0,014) terhadap BT dan tidak terdapat interaksi antara grup dan parameter bobot tubuh linear berdasarkan analisis kovarian. Tingkat mortalitas Kambro lebih rendah dibandingkan $F_1$ Pelung tanpa vaksinasi dengan sistem pemeliharaan semi-intensif. Sebagai akibat dari ukuran populasi hibrida terbatas, temuan riset harus divalidasi dengan ukuran populasi hibrida lebih besar.

Kata kunci: Broiler Cobb 500; *grandparent stock*; Kambro; Pelung Blirik Hitam; persilangan selektif.

## INTRODUCTION

Pusat Data dan Sistem Informasi Pertanian (2015) stated that chicken meat consumption rate of 2014 reached 4.48 kg/capita/year (total consumption of broiler chicken, post-laying layer chicken and male layer also native chicken). Chicken poultry sector contributed around 60.73% of the demand on animal consumption needs fulfillment (Suprijatna, 2010). Ditjen PKH (2017) showed that native chicken production nationally reached 8.50 % or 284.9 thousand tons with contribution percentage of 12.86 % to nation chicken meat production. Ditjen PKH (2018) showed that Indonesia poultry livestock populations in 2018 consisted of 1.8 billion broiler-type/broiler chickens, 181.752 layer chickens and 310.960 native chickens. Broiler-type and laying-type chicken poultry industry went through significant growth per year driven by improvement on income and knowledge on healthy nutritional-balance food product (Iskandar, 2017). Chicken poultry industry in Indonesia is still depending on imported broiler caused by short production period and rapid turnover (Nurfadillah *et al.*, 2018).

Native chicken has unlimited potential to become broiler-type, egg-type and dual purpose chicken candidate in order to fulfill domestic consumption needs of animal based food through selective and genetic engineering (Nataamijaya, 2010; Henuk and Bakti, 2018; Kartika *et al.*, 2016). Native Indonesia chickens are called *Kampung* chickens or native (non-breed chickens) to differentiate commercial breed such as Cobb, Hubbar, Hybro, Isa Hyline and Hisex (Henuk and Bakti, 2018). Identification of native chicken germplasm resulted in 34 breeds of chicken consist of *Ayunai, Balenggek, Banten, Bangkok, Burgo, Bekisar, Cangehgar, Cemani, Ciparage, Gaok, Jepun, Kampung, Kasintu, Kedu (hitam* and *putih), Pelung, Lamba, Maleo, Melayu, Merawang, Nagrak, Nunukan, Nusa Penida, Olagan, Rintit atau Walik, Sedayu, Sentul, Siem, Sumatera, Tolaki, Tukung, Wareng, Sabu,* and *Semau* (Henuk and Bakti, 2018). Approximately 11 native chicken breeds are categorized as candidates of broiler-type and laying-type chicken (Henuk and Bakti, 2018). Native Indonesia chickens have to be maintained optimally in order to support small scale poultry industry based on native chickens. Native Indonesia chickens germplasm can be the solution for fulfilling the increasingly domestic food consumption demand (Daryono *et al.*, 2010). Ningsih and Prabowo (2017) stated that various challenges faced by poultry industry sub sector especially broiler, besides market integration several factors including production, productivity and competitiveness of poultry product. Nurfadillah *et al*. (2018) stated that agribusiness problem in subsystem of broiler chicken poultry is economy efficiency in poultry level added by high cost production inflicted by dependence on imported raw-material of feed. Improvement of efficiency and poultry product quality are decided by supply of superior chicken breed, feed demand fulfillment and good rearing management system (Anggitasari *et al.*, 2016). Improvement of productivity and competitive quality of local broiler chicken can be achieved through selective breeding of native Indonesia chicken breeds. Selective breeding is aimed to produce superior chicken breed with adjusted phenotype quality based on human needs (Das *et al.*,2008; Cheng, 2010; Oldenbroek and van der waaij, 2014; Mariandayani *et al.*, 2017; Sudrajat and Isyanto, 2018).

*Pelung Blirik Hitam* has several distinguished characters such as posture and body weight compare with otther native breeds (Daryono *et al*., 2010). Body weight of male *Pelung* chicken can reach 3.37 kg and female can reach 2.52 kg (Daryono *et al*., 2010). Broiler Cobb 500 has distinguished productivity and high growing rate in grower phase (7 to 18 weeks). Male and female Broiler Cobb 500 can reach 1,599.17 grams and 1,540.46 gram (Hassan *et al.*, 2016). This research is aimed to measure the phenotypic performance of hybrid chicken *Kambro* based on research conducted by Tamzil *et al*. (2018) to *Cairina moschata* and Daryono *et al*. (2010) to hybrids from crossbreeds of *Pelung* with *Cemani* with several addition and adaptation of measurement parameters. Measurement parameters used in this research are estimation model of body weight, body weight growth, linear body weight parameter, mortality rate, phenotypes and vitality parameter. Empowerment of native Indonesia chicken can contribute to availibity of food source and support native Indonesia chicken germplasm conservation (Suprijatna, 2010; Sudrajat and Isyanto, 2018).

## RESEARCH METHODS

This research was conducted in Pusat Inovasi Agroteknologi (PIAT), Desa Kali Tirto, Berbah, Sleman Regency, Yogyakarta using 4 females Broiler Cobb 500 and 1 male *Pelung Blirik Hitam*, 1 female *Pelung Blirik Hitam*, 7 $F_1$ *Pelung* chickens, 22 Broiler Cobb 500 chickens and 17 *Kambro* ($F_1$ Broiler) chickens. $F_1$ *Pelung* was produced from crossbreeds of *Pelung Blirik Hitam* native to Cianjur, West Java (Fig. 1B). Broiler Cobb 500 was produced by rearing of Day Old Chicken (DOC) Broiler Cobb 500 from Pokphand Indonesia. *Kambro* ($F_1$ Broiler) was produced from crossbreeds of 4 female Broiler Cobb 500 aged 6 months with a male *Pelung Blirik Hitam* (Fig. 1A). Parental crossbreeding was conducted in broodshed (8 $m^2$) owned by Gama Ayam Research Team. Standard feeds produced by PT. Japfa Comfeed Indonesia, AD II (brood/juvenile, 9-22 week) and BR-1 (starter, 0-22 day) with *ad libitum* dietary. Supplemental vitamin *Egg Stimulant®* and *TetraChlor®* produce by Medion was needed to improve immunity and brood productivity. Egg collection from each crossbreed was hatched using incubator. Day Old Chicken (DOC) was reared intensively in bamboo pens insulated by plywood and equipped with incandescent lamps (15 watts). Chicken aged 4 week then transferred into larger shed with semi-intensive rearing system (8 $m^2$). Grouping of each chicken based on its crossbreeding as follow DOC $F_1$ *Pelung* (group I), DOC Broiler Cobb 500 (group II) and DOC *Kambro* (group III). Body weight growth of DOC Broiler, DOC $F_1$ *Pelung* and DOC *Kambro* was measured per week with digital scale *KrisChef EK9350H* with 0.01 gram accuracy until chicken reach 8-weeks-old. Zoometrical measurement was measured with metline based on morphological guidance of chicken sceletal (Supplemental Files, adapted with modification and addition from Daryono *et al.*, 2010).

1. TA was measured from the digit/hallux to the tip of the comb
2. TB was measured from the digit/hallux to the end of the distal vertebrae
3. LP was measured from articular to dexter
4. PP was measured from the base of the angular process to the end of the mandibular symphysis
5. PK was measured from the supraorbital bone to premaxilla
6. LK was measured from quadratojugal sinister to dexter
7. TJ was measured from the highest tip of the comb to the base of the comb
8. PJ was measured from the back to the front of the comb
9. PB was measured from the tip of the first thoracic vertebra to the base of the pygostyle
10. LB was measured from the base of the femoral bone to dexter
11. LD was measured from the sternal of the keel in a circle
12. PPu was measured from the thoracic vertebrae to the caudal vertebrae end
13. PS was measured from the base of the humerus to the end of the carpus
14. PL was measured from the base of the atlas to the tip of the thoracic vertebrae
15. PBe was measured from the tip of the femur to the base of the tibiotarsus
16. PPa was measured from the end of the patella to the base of the femur

Linear body weight parameter consists of TA (chicken height), TB (body height), PB (body length), LB (body width), PPu (dorsal length), PL (neck length), PS (wingspan) amd LD (chest circumference). Vitality parameter consists of TJ (comb height), PJ (comb length), PK (head length), LK (head width), PP (beak length) and LP (beak width). Qualitative phenotype parameter including neck feather colour, dorsal/ back feather colour, chest feather colour, body feather colour, femoral feather colour, shank colour, comb colour, comb shape and beak colour. Phenotype parameter of hybrid is identified as visual data with black background photo. Weekly data record consists of body weight growth (BT) and femur-tibia length (PPa-PBe). Data is analyzed with correlation, regression, one way anova and independent sample t-test using IBM© SPSS© *Statistics version* 21. Independent sample t-test can be used to compare average body weight, body growth, feed intake, feed conversion and mortality rate between two chicken populations (Darwati *et al.,* 2016). Correlation between femur-tibia length and linear body weight parameter to body weight growth are analyzed with Pearson correlation method, linear regression, multiple linear regression and Analysis of Covariance (ANCOVA). Phenotype parameter is analyzed with visual observation scoring method based on photo.

## RESULTS AND DISCUSSION

Crossbreeds of female Broiler Cobb 500 with male *Pelung Blirik Hitam* produced 18 hybrids named *Kambro* consisted of 9 males *Kambro* and 9 females *Kambro* (Fig. 1A$_{1-2}$). Crossbreeds of female *Pelung Blirik Hitam* with male *Pelung Blirik Hitam* produced 22 F$_1$ *Pelung* chickens (Fig. 1B$_{1-2}$).

Day Old Chicken (DOC) of control populations consist of F$_1$ *Pelung* and Broiler Cobb 500 each with 22 chickens. Mortality rate of group III (*Kambro*), group II (Broiler Cobb 500) and group I (F$_1$ *Pelung*) subsequently are 5.5%, 0% and 68.2%. Mortality rate of group I is higher than group III and II. Earliest record of death was in group I at 2-weeks-old meanwhile in group III at 6-weeks-old. Most probable cause of death in group I and group III caused by infection of infectious coryza (snot) through daily observation. Infectious coryza (snot) disease is caused by gram-negative bacteria *Haemophilus paragallinarum* with symptom of rapid infection and high morbidity, declining in egg production, oculonasal conjunctivitis, face swelling and conjuncivital sac exudation (Ali *et al.,* 2013; Iskandar, 2017).

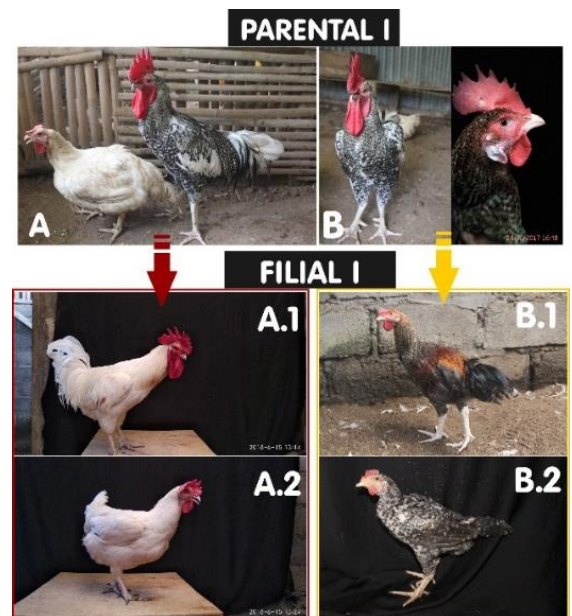

Figure 1. Chicken cross diagram. Parental I (A: female Broiler Cobb 500 and male *Pelung Blirik Hitam*; B: female *Pelung Blirik Hitam* and male *Pelung Blirik Hitam*) and Filial I (A.1: male *Kambro*; A.2: female *Kambro*; B.1: male F$_1$ *Pelung*; B.2: female F$_1$ *Pelung*) (Personal Documentation, 2017)

The absence of vaccination was a treatment to assess immunity of each chicken groups. Data of mortality from each groups lead to a conclusion that immunity resistance of group III was higher than group I. High mortality rate can be caused by the absence of vaccination in group I and III. Group II had been vaccinated since hatch by DOC producer. *Kampung* chicken has distinguished immunity resistance better than other native tropical broiler-type breed and highest expression of antivirus gene *Mx+* (Diwyanto and Prijono, 2007; Nuroso, 2010; Kartika *et al.,* 2016; Nurhuda, 2017). *Kambro* has higher immunity resistance indicates improvement of genetic quality of native chicken through crossbreed and semi-intensive rearing system supported by several management and environmental factors.

Egg collection and hatching of *Kambro* was 10 until 20 eggs per week during 6 months period (December 2017 until May 2018). Egg productivity rate was low at 20 until 22 eggs on peak of Broiler Cobb 500 laying cycle. Female Broiler Cobb 500 (±6 months) hatchability only reach 25% per hatching period. Several factors were influencing the fluctuation of *Kambro's* egg productivity including nutrition, stress level, sperm fertility and egg fertility. Laying broiler productivity reaches its peak of laying cycle at the age of 23 weeks (± 6 months) (Rahman *et al.*, 2015). Female Broiler Cobb 500 egg productivity in this research can be influenced by female age. Hameed *et al.* (2016) stated that egg weight and hatchability can be influenced by female aging, declining hatchability of eggs reach 15% in female broiler at 30-weeks-old with egg weight less than 60 grams. Main factor that influenced the fluctuation of egg productivity can be caused by *ad libitum* standard feed dietary. Rahman *et al.* (2015) stated that *ad libitum* dietary can decrease egg productivity, minimizing egg hatchability and increasing mortality rate. Feed diet restriction must be implemented to limiting body weight growth, maximizing egg production and increase the female Broiler Cobb 500 fertility (Rahman *et al.*, 2015).

In the table 1. are shown the results of one way anova analysis of PPa, PBe and BT on each chicken groups which ssignificantly different (p<0.01). BT shows a highly significant (p<0.01) difference in three groups of chicken [$F_{(2, 43)} = 62.09$, p<0.01, $\eta^2 = 0.743$]. PPa shows a highly significant (p<0.01) difference in three groups of chicken [$F_{(2, 43)} = 55.09$, p<0.01, $\eta^2 = 0.719$]. PBe shows a highly significant (p<0.01) difference in three groups of chicken [$F_{(2, 43)} = 22.87$, p<0.01, $\eta^2 = 0.515$]. Post hoc analysis with Fisher's LSD indicates a significant difference of PPa, PBe and BT on each of chicken groups. PPa of group I (M = 6.79, SD = 1.03) significant to group II (M = 5.69, SD = 0.82) and group III (M = 8.92, SD = 1.08). PBe of group I (M = 8.9, SD = 0.82) significant to group II (M = 11.25, SD = 0.85) and group III (M = 11.96, SD = 1.2). BT of group I (M = 602.88, SD = 79.93) significant to group II (M = 1,706.82, Sd = 262.54) and group III (M = 1,244.14, SD = 453.82). Conclusively group III shows a distinguished performance of BT, PPa and PBe compare with group I (Fig. 2A). BT of group III (1,244.14 ± 453.82 gram) approaches BT of group II (1,706.82 ± 262.54 gram) at 8-weeks-old. One way anova analysis of PPa, PBe and BT is strengthen with independent sample t-test (Supplemental File 3, Table 2). PPa of group III (M = 8.92, SD = 1.08) is significant to group I (M = 6.79, SD = 1.03), $t_{(22)} = 4.446$, p<.001. PPa of group III (M = 8.92, SD = 1.08) is significant to group II (M = 6.79, SD = 1.03), $t_{(37)} = 10.62$, p<.001. PBe of group III (M = 8.92, SD = 1.08) is significant to group I (M = 6.79, SD = 1.03), $t_{(22)} = 5.956$, p<0.01. PBe of group III (M = 8.92, SD = 1.08) is significant to group I (M = 6.79, SD = 1.03), $t_{(37)} = 2.139$, p<0.05. BT of group III (M = 8.92, SD = 1.08) is significant to group I (M = 6.79, SD = 1.03), $t_{(21,66)} = 9.88$, p<0.01. Variance test with Levene's test of BT group III-I indicates a dissimilarity (F = 11.11, *p* = 0.003), as adjustment the degree of freedom is set from 22 into 21.66.

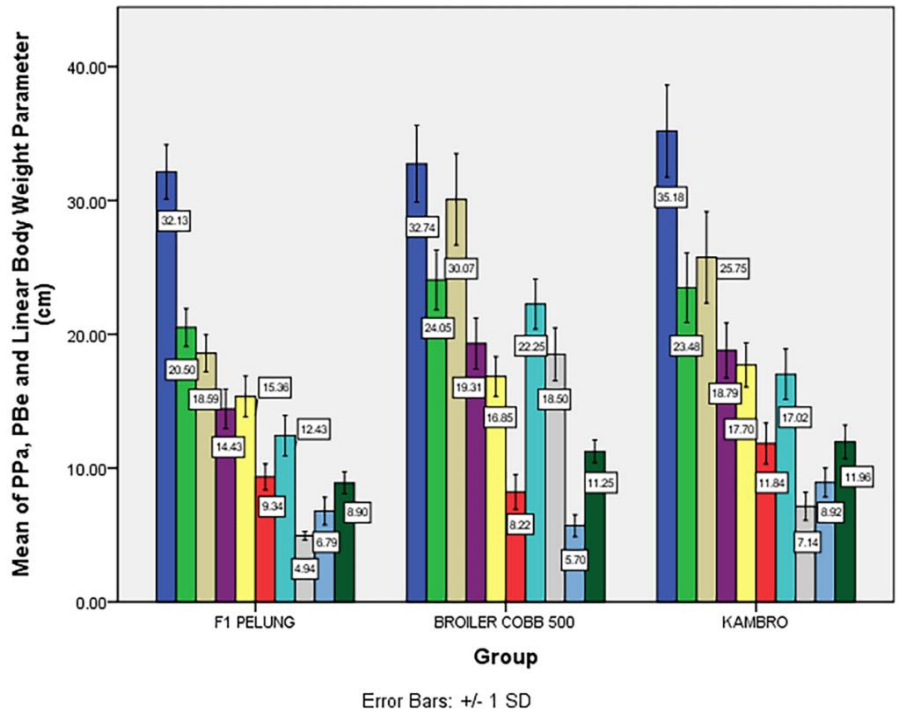

(A)

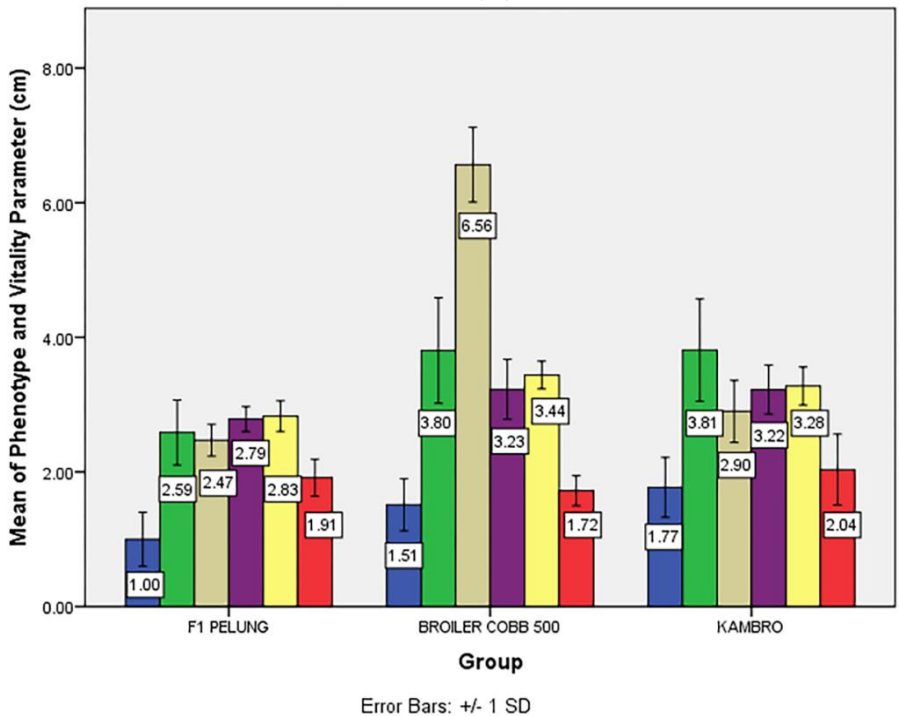

(B)

Figure 2. (A) Mean of PPa, PBe and linear body weight parameters of group I, II and III in 8 weeks; (B) The mean parameters of the chicken group I, II and III vitality and phenotype in 8 weeks. The standard deviation is denoted by T-bar. In graph A each parameter is symbolized by arrangement: **TA** ▪; **TB** ▪; **PB** ▪; **LB** ▪; **LD** ▪; **PPu** ▪; **PS** ▪; **PL** ▪; **PBe** ▪; **PPa** ▪. In graph B each parameter is symbolized by arrangement: **LP** ▪; **PP** ▪; **PK** ▪; **LK** ▪; **TJ** ▪; **PJ** ▪

Table 1. Analysis of One Way Anova PPa, PBe and BT Chicken Groups I, II and III at 8-weeks

| Parameters | Chicken Groups | | | F | η2 |
|---|---|---|---|---|---|
| | I (n = 7) | II (n = 22) | III (n =17) | | |
| PPa (cm) | 6.79a (1.03) | 5.69b (0.82) | 8.92ab (1.08) | 55.09*** | 0.719 |
| PBe (cm) | 8.9a (0.82) | 11.25b (0.85) | 11.96ab (1.2) | 22.87*** | 0.515 |
| BT (gram) | 602.88a (79.93) | 1,706.82b (262.54) | 1,244.14ab (453.82) | 62.09*** | 0.743 |

PPa = Femur Length  
PBe = Tibia Length  
BT = Body Weight  
\* = p <0.05; \*\*\* = p <0.01. The standard deviation is listed below the mean. The averages with different subscripts in the same column differ significantly (p<0.05) based on Fisher's LSD post hoc.

Average Body Weight (BT) of *Kambro* at 8-weeks-old can be compared with other similar crossbreed. Crossbreeding of *Sentul* chicken reached average body weight of 896.34 ± 55.46 grams (male *Sentul*) and 736.00 ± 46.63 grams (female Sentul) during 75 days period (Solikin *et al*., 2016; Sudrajat and Isyanto, 2018). Mariandayani *et al*. (2013) stated data about several body weight of native chicken at 8-weeks-old which including *Pelung* (male 458.23 grams and female 420.11 grams), *Sentul* (male 406.36 grams and female 355.98 grams), *Kampung* (male 411.56 grams and female 358.74 grams). From this comparisons can be concluded that *Kambro* can reach higher BT than othher native chicken breeds. Hasyim (2015) stated that hybrids chicken crossbreeds of *Kampung* and Broiler at 12-weeks-old can reach 2,335 grams (male) dan 1,833 grams (female). *Kambro* body weight growth in the 8 weeks of age has not reach inflection point whereas *Kambro* BT growth projection was estimated to be higher as weeks follow. Inflection point is maximum body weight growth, during this period a shift of growth phase occurs with declining growth. Growth can occur during weeks follow because chicken has not reached sexual maturity (Sogindor, 2017). Suprijatna (2010) stated that sexual maturity of *Pelung* chicken at day-165 with 12-weeks-old weight can reach 669 grams. Nurhuda (2017) stated that genetic component combination affects BT of chicken from crossbreeding with hybrids observed to have better performance than parantal generation on several characters or traits. Average BT of *Kambro* was 1,244.14 ± 453.82 grams lower than Broiler Cobb 500 which can reach 1,706.82 ± 262.54 grams at 8-weeks-old for the reason of only inherited 50% of Broiler Cobb 500 genetic components, whereas BT of $F_1$ *Pelung* only reached 602, 88 ± 79,93 grams with the same period.

PPa, PBe and several linear body weight parameters have correlations with chicken body weight (Ukwu *et al.,* 2014). Linear body weight parameter consist of shank length, chest circumference, tibia length, neck length, dorsal length and femur length (Ukwu *et al.,* 2014). Linear body weight parameter used in this research consist of TA, TB, LB, PL, PS, LD and PPu. Linear body weight parameter has significant influence on selective breeding program, also as chicken body weight indicator and market attraction (Ukwu *et al.,* 2014; Assan, 2015). LD and PB in group III showed significant (p<0.01) results than group II, meanwhile TA, PL and LB in group III were superior than group II (Table 2). Performance improvement of *Kambro* to *Pelung* was shown by significant results of linear body weight parameter in group III and group I. PPa, PBe and linear body weight parameter correlation to BT is summarized in Table 2.

Table 2. Correlation of linear body weight parameter, PPa and PBe to BT in chicken group I, II and III

| Parameters (cm) | | Chicken Groups | | |
|---|---|---|---|---|
| | | I (n=7) | II (n=22) | III (n=17) |
| Linear Body Weight Parameters | TA | -0.374$^{ns}$ | 0.444$^*$ | 0.553$^*$ |
| | TB | -0.091$^{ns}$ | 0.380$^†$ | 0.633$^{**}$ |
| | LB | 0.344$^{ns}$ | 0.216$^{ns}$ | 0.629$^{**}$ |
| | PB | 0.150$^{ns}$ | 0.005$^{ns}$ | 0.478$^†$ |
| | PL | -0.454$^{ns}$ | 0.361$^†$ | 0.152$^{ns}$ |
| | PS | 0.792$^*$ | 0.179$^{ns}$ | 0.606$^{**}$ |
| | LD | 0.131$^{ns}$ | -0.398$^†$ | 0.396$^{ns}$ |
| | PPu | 0.431$^{ns}$ | 0.349$^{ns}$ | 0.299$^{ns}$ |
| PPa | | 0.975$^{***}$ | 0.932$^{***}$ | 0.965$^{***}$ |
| PBe | | 0.298$^{ns}$ | -0.064$^{ns}$ | 0.567$^*$ |

† = $p<0.10$, * = $p<0.05$, ** = $p<0.01$, *** = $p<0.001$. † *very slightly significant*, ns = *non-significant*

Table 3. Chicken group X linear body weight parameter factor (FAC1_1) ANCOVA body weight (BT) at 8 weeks

| Source | *Df* | *F* | η2 | *p* |
|---|---|---|---|---|
| Group | 1 | 7.205 | 0.265 | 0.014 |
| FAC1_1 | 1 | 2.508 | 0.111 | 0.129 |
| Group* FAC1_1 | 1 | 0.482 | 0.024 | 0.482 |
| Error (within groups) | 20 | | | |

FAC1_1: TA, TB, LB, LD, PL, PS, PPu dan PB; $p<.05$

In Table 2. Pearson's correlation analysis indicated significant positive correlation between PPa and PBe to BT group III (PPa *r* (17) = 0.965, *p*<0.01; PBe *r* (17) = 0.567, *p*<0.01). In group I and group II, BT only has positive correlation with PPa (group I *r* (7) = 0.975, *p*<0.01; group II *r* (22) = 0.932, *p*<0.01). In group III TA (0.553), TB (0.633), LB (0.629) and PS (0.606) significantly correlates (p<0.05) with BT. In group II TA (0.444) significantly correlates (p<0.05) with BT. In group I PS (0.792) positively correlates (P<0.05) with BT. Linear body weight parameter on each groups have weak positive correlation to BT meanwhile PPa have strong positive correlation to BT (r > 0.90) within each groups. Conclusively PPa can be used as standardized BT estimation model within all chicken groups. Regression analysis was used to strengthen this conclusion as can be seen in Figure 3.

Non-linear regression model applied in this research was curvilinear quadratic because of improvement on $R^2$ value compare to $R^2$ in simple linear regression (Supplemental File 3, Table 4). PPa is contruction parameter of prediction model suitable in non-linear projection of BT in chicken group I, II and III. BT prediction model according to linear body weight parameter with positive weak correlation analyzed with ANCOVA in Table 3.

ANCOVA annalysis between subject and factor [Chicken Group (I, II, and III); covariate: FAC1_1] showed significant group effect F (1.20) = 7.205, p = 0.014, η2 = 0.265, while FAC1_1, F (1, 20) = 2.508, p = 0.129, η2 = 0.111 was insignificant, and no interaction between group and FAC1_1, F (1.20) = 0.482, p = 0.482, η2 = 0.024. ANCOVA analysis strengthening PPa parameter as 8 weeks age *Kambro* Body Weight (BT) prediction model. Semakula *et al.*, (2011) stated that native Lake Victoria chickens live body weight correlates with chest girth. On that research live body weight prediction model and chest girth is non-linear regression highest $R^2$ value on power model ($0.001G^{2.417}$) (Semakula *et al.*, 2011). Ukwu *et al.*, (2014) stated that linear body weight parameter including shank length can be used as live body weight prediction model of native Nigeria chicken. Mabelebele *et al.* (2017) stated that Broiler Ross 308 has distinguished femur and tibia length compare to Venda chicken, native South Africa chicken. Similar phenomenon can be observed in *Pelung* chicken with shorter PPa compare to Broiler Cobb 500, on the other hand distinguished in PBe. *Kambro* chicken has PPa and PBe superior than parental generation (Fig. 2A). Mabelebele *et al.* (2017) stated that polinomial regression of Ross 308 carcass weight was inflicted 97% by femur length and 94% by tibia length, meanwhile Venda chicken was inflicted 89% by tibia length and 37% by femur length. Dalam penelitian ini regresi non-linear quadratic fungsi bobot tubuh ayam Broiler Cobb 500 dipengaruhi oleh 97,8% PPa, ayam $F_1$ *Pelung* 96,2% PPa dan ayam *Kambro* 95,6% PPa. Conclusively PPa length growth coherently following BT growth in *Kambro*.

PPa function of Broiler Cobb 500 was higher than $F_1$ *Pelung* can be caused by intensive rearing system. *Pelung* chicken mostly reared with extensive system or free-range with variative feed diet, $F_1$ *Pelung* bone growth retardation can be caused by locomotion limitations. Henuk and Bakti (2018) stated that extensive rearing system decrease native Indonesia chicken productivity because feed diet inefficiency and lengthy growth period 90 day/1kg. Femur length (PPa) growth adjusted with Body Weight (BT) in Broiler Cobb 500 with extensive rearing system and non-strict feed diet impacting negatively growth performance of fast-growing broiler chickens (Pauwels *et al.*, 2015). Regression analysis of group II PPa parameter showed declining BT by increasing length of PPa (Fig. 3). Broiler locomotion was affected by BT and PPa. Strict diet can cause muscoscceletal growth delay in broiler with further implication muscle stress of movement and locomotion (Paxton *et al*., 014). Shim *et al*. (2012) stated that bone of fast-growing broiler at 6 weeks age is longer, wider, heavier, stronger, compact and high calcium concentrated compared with slow-growing broiler with the same age. Han *et al*. (2015) stated that tibia is the longest and heaviest part compared with femur as the longest diameter bone. Mortality rate and performance of broiler-type chicken are affected by bone structure. Bone growth abnormality can be affected by several factors including lighting period. Van der Pol *et al*. (2015) stated that minimum lighting period decreases environmental stress of chickens, where extreme dim-bright lighting increase asymetric bone growth in broiler. PPa function of *Kambro* was lower than $F_1$ *Pelung* and Broiler Cobb 500 conclusively semi-intensive rearing system and non-strict combination of feed diet can be standardized as suitable *Kambro* rearing system.

Market assessments and crossbreed selection depend on visual phenotype parameter appearance (Frame, 2009; Semakula *et al*., 2011; Assan, 2015). Visual method can be used to rapidly identify certain traits quality of chickens.

Navara *et al*. (2012) stated that phenotype appearance determines chickens genetic succession and productivity. Based on LP group III was insignificant (p>0.05) to group I (Table 2). Based on PJ, TJ and LK group III was insignificant (p>0.05) to group II (Fig. 2). Comb colour of group III was dominated 58.82% by red colour and 41.18% rosy colour (Table 4) with 100% of single shaped comb. Navara *et al*. (2012) stated that comb colour has positive significant correlation to sperm function, on the other hand comb size has negative significant correlation. These findings was contradictive with other findings which stated that comb size has positive significant correlation to vitality, sperm function and mating signal in male (Gebriel *et al.,* 2009; El Ghany *et al*., 2011; Udeh *et* *al.*, 2011). Dominant male showed larger comb dimention (PJ+TJ) with bright red colour with low sperm motility (Navara *et al.*, 2012). Female inclination to select dominant male can cause quality reduction of filial generation sperm quality (Navara *et al.*, 2012). Frame (2009) stated that comb colour involves as an indicator of laying period with pale coloured indicates laying initiation and post-laying period while bright red coloured indicates optimum laying period. Average PJ of *Kambro* is 3.81 ± 0.76 cm shorter than several other breed such as White Leghorn (10-16 cm), Red Junglefowl (6-12 cm) and broiler (8-14 cm) (Navara *et al.*, 2012).

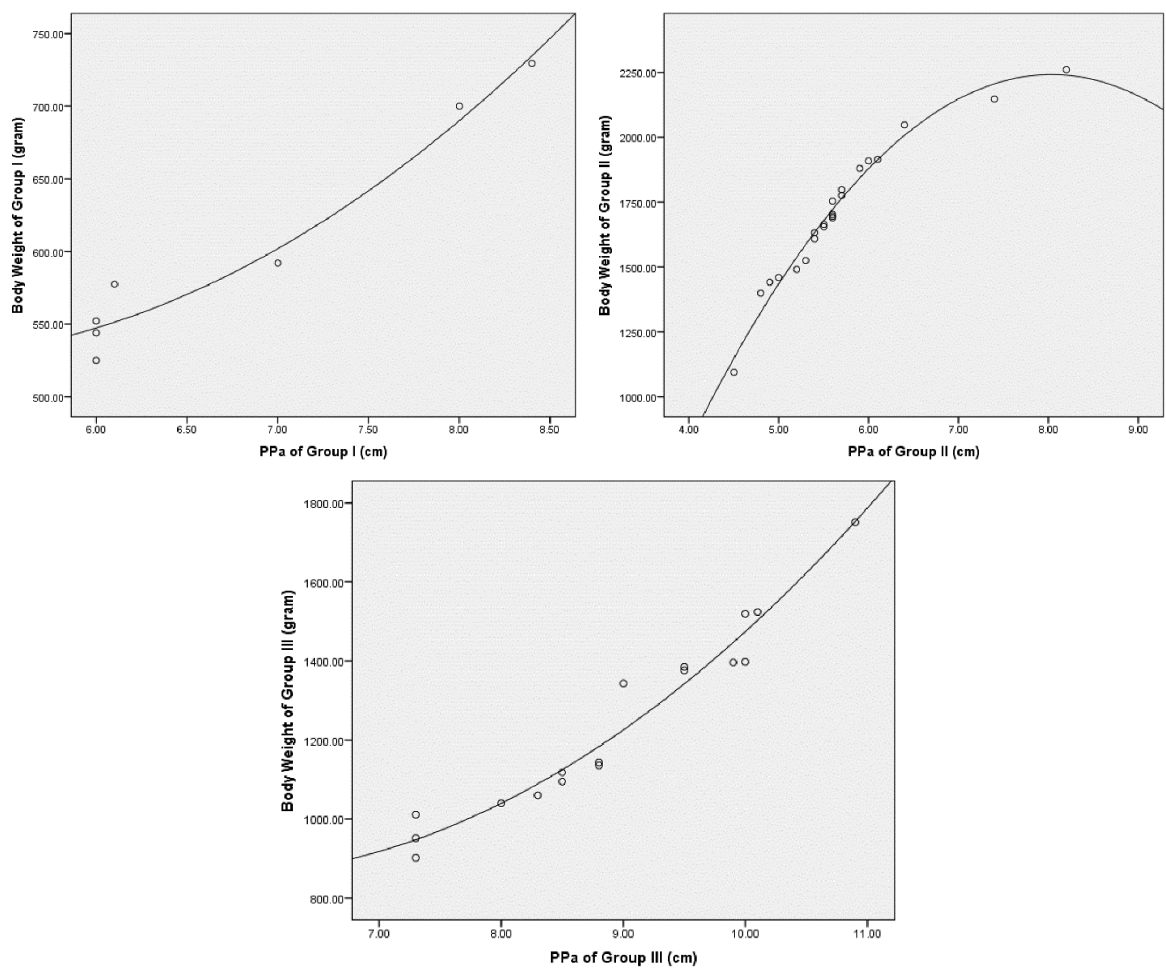

Figure 3. Curvilinear quadratic PPa model against BT chicken group I, II and III.

Identification of comb colour, PJ and TJ in *Kambro* become parental selection guide, in order to sorting out male with low sperm motility in future crossbreed. Based on this findings in the next crossbreed male with small PJ and TJ and female with bright red colour will be chosen. Measurement on PK, LK, PP and LP (Fig. 2B) is an indicator of dietary pattern and feed consumption rate with correlative link between these parameter and Body Weight (BT) have been clarified by several studies (Joller *et al.*, 2018; Fayeye *et al.*, 2013; Fahey *et al.*, 2007; Yakubu *et al.*, 2009). Beak deformity has known to affecting dietary pattern and chickens body weight (Joller *et al.*, 2018) with genetic influence of *DEGs* gene expression (Bai *et al.*, 2014). Beak colour was dominated by ivory white colour (70.58%) followed by white black patterned (29.42%). Frame (2009) stated that fading of beak colour from white into dull white or ivory white colour indicate chicken age between 4 to 6 weeks old. Chicken shank colour controlled by locus allele *Id-id* and *W-w* with *Id-* expressing white or yellow colour and *idid* expressing black, gray or green colours influenced by *GRAMD3* gene in dermal tissue of shank (Xu *et al.,* 2017). Frame (2009) stated that depigmentation of shank colour is an indication of hen egg productivity during 15 to 20 weeks.

Feather expression in *Pelung Blirik Hitam* is determined by three genotypes $Z^BZ^b$ (patterned), $Z^bZ^b$ (plain) and Broiler Cobb 500 is determined by genotype $Z^bW$ (plain white). Hasnelly *et al.* (2017) stated that chicken feather colour is expressed by (s) allele for old golden colour, (S) for silver recessive, (b) for plain colour, (B) for patterned colour. Feather colour in broiler chicken can be classified into dominant white which can be observed in White Leghorn chicken with several variation that are smoky / grayish (*I\*S*) and dun / whitish (*I\*D/i*) (Kerje *et al.*, 2004). Both male *Kambro* ($Z^BZ^b$) and female *Kambro* ($Z^BW$) showed 100% gene frequencies of feather colour with black pattern, brown, and gray. *Pelung Blirik Hitam* and Broiler Cobb 500 shank colour genotype can be categorized as *IdId*/*Id_* (white/yellow) and *idid* (black/gray/green). Both male *Kambro* and female *Kambro* shank colour gene frequencies can be classified into three groups *IdId*/white (52.95%), *Idid*/white with black pattern/gray (41.17%) and *idid*/black (5.88%). Body feather colour and shank colour variation of *Kambro* indicate segregation of alleles in the population which inherited by *Pelung Blirik Hitam* and Broiler Cobb 500. Duguma (2006) stated that bright or white body feather colour has higher commercial value and qualified by market. Semakula *et al.* (2011) stated that visual judgement has significant influence on sale value with a tendency of increasing demang of native Ugandan chickens. Suprijatna (2010) stated that native Indonesia chickens has a *niche* market and the prevalencies of customer showed a higher demand on native chickens based of its unique taste and phenotypic appearance.

Table 4. Phenotype parameter of *Kambro* at 8 weeks based of visual observation scoring method.

| Phenotype Parameters | Characters | Gene Frequency (%) ♂/♀ (n=17) | Locus | Gene |
|---|---|---|---|---|
| Color of neck feather | White | 100 | *I-i* | $q^I$-$q^i$ |
| Back feather color | White with black, brown and gray strands | 100 | *I-i/ E-e+-e* | $q^I$-$q^i/q^E$-$q^{e+}$-$q^e$ |
| Color of chest hair | White | 100 | *I-i* | $q^I$-$q^i$ |
| Body feather color | White with black, brown and gray strands | 100 | *I-i/ E-e+-e/B-b* | $q^I$-$q^i/ q^E$-$q^{e+}$-$q^e/q^B$-$q^b$ |
| Color of femoral feather | White | 52.95 | *I-i* | $q^I$- |
|  | White black or gray pattern | 47.05 | *E-e+-e/B-b* | $q^E$-$q^{e+}$-$q^e/q^B$-$q^b$ |
| Shank color | White | 52.95 | *Id- id* | $q^{Id}/q^{id}$ |
|  | White black or gray pattern | 41.17 | *Id- id* | $q^{Id}/q^{id}$ |
|  | Black | 5.88 | *Id- id* | $q^{Id}/q^{id}$ |
| Comb color | Red | 58.82 | - | - |
|  | Pink | 41.18 | - | - |
| Comb shape | Single | 100 | *P-p* | $q^P/q^p$ |
| Beak color | Broken white | 70.58 | - | - |
|  | White black pattern | 29.42 | - | - |

## CONCLUSION

Based on measurement of Body Weight (BT), *Kambro* (1244.14 ± 453.82 grams) performed significantly (p<0.01) better than $F_1$ *Pelung* (602.88 ± 79.93 grams) in 8 weeks period with *ad libitum* diet of standard feed. The performance escalation of *Kambro* compared with $F_1$ *Pelung* was significant based on the measurement of linear body weight parameter, vitality parameter, femur length (PPa), tibia length (PBe) and phenotype parameter. Observation of phenotype parameter showed the resemblance of *Kambro* with parental generation. Estimation model of *Kambro* Body Weight (BT) can be measured with femur length (PPa) in non-linear quadratic regression (r = 0.956) based on this formula $1.84E3 \pm 3.54E2*x+31.73*x^2$. ANCOVA analysis showed no interaction between group and linear body weight parameter and there was significant difference BT of groups (p =0.014). Mortality rate of *Kambro* was lower than $F_1$ *Pelung* with the absence of vaccination in semi-intensive rearing system. Research findings must be validated with larger population size.

## SUGGESTION

Further research with larger number of hybrid chickens must be conducted to validate the result on this study.

## ACKNOWLEDGEMENT

Research funding of Ministry of Higher Education Republic of Indonesia. Gama Ayam Research Team and Agrotechnology Innovation Center Universitas Gadjah Mada.

Buletin Veteriner Udayana    Volume 11 No. 2: 188-202
pISSN: 2085-2495; eISSN: 2477-2712    Agustus 2019
Online pada: http://ojs.unud.ac.id/index.php/buletinvet    DOI: 10.24843/bulvet.2019.v11.i02.p12http://ditjenpkh.pertanian.go.id/userfiles/File/Buku_Statistik_2017_(ebook).pdf?time=150512744 3012

Direktorat Jenderal Peternakan dan Kesehatan Hewan. 2018. Statistik Peternakan dan Kesehatan Hewan 2018 *Livestock and Animal Health Statistics* 2018. [accessed: April 29th 2019]. http://ditjenpkh.pertanian.go.id/userfiles/File/Buku_Statistik_2017_(ebook).pdf?time=150512744 3012

Diwyanto K, Prijono SN. 2007. Keanekaragaman sumber daya hayati ayam lokal Indonesia: manfaat dan potensi. Bogor (Indones): Lembaga Ilmu Pengetahuan Indonesia.

Duguma R. 2006. Phenotypic characterization of some indigenous chicken ecotypes of Ethiopia. *Livestock Res. Rural Dev*. 18(9): 131.

El Ghany FAA, El Dein A, Soliman MM, Rezaa AM, El Sodany SM. 2011. Relationships between some body measurements and fertility in males of two local strains of chicken. *Egypt Poult. Sci*. 31:331-349.

Fahey AG, Marchant-Forde RM, Cheng HW. 2007. Relationship between body weight and beak characteristics in one-day-old White Leghorn chicks: Its Implications for Beak Trimming. *Poult. Sci*. 86: 1312-1315.

Fayeye TR, Hagan JK, Obadare AR. 2013. Morphometric traits and correlation between body weight and body size traits in Isa Brown and Ilorin ecotype chickens. *Iranian J. Appl. Anim. Sci*. 4(3): 609-614.

Gebriel GM, Kalamah MA, El-Fiky AA, Ali AFA. 2009. Some factors affecting semen quality traits in norfa cocks. *Egypt Poult. Sci*. 29: 677–693.

Hameed T, Mustafa MZ, Taj MK, Asadullah, Bajwa MA, Bukhar FA, Kiani MMT, Ahmed A. 2016. Hatchability and fertility in broiler breeder stock. *J. Chem. Biol. Phy.Sci*. 6(2): 266-274.

Han JC, Qu HX, Wang JG, Chen GH, Yan YP, Zhang JL, Hu FM, You LY, Cheng YH. 2015. Comparison of the growth and mineralization of the femur, tibia, and metatarsus of broiler chicks. *Braz. J. Poult. Sci*. 17(3): 333-340.

Hasnelly, Iskandar S, Sartika T. 2017. Karakteristik kualitatif dan kuantitatif ayam SenSi-1 Agrinak. *JITV*. 22(2): 68-79.

Hassan KMd, Kabir HMd, Sultana S, Hossen AMd, Haq MM. 2016. Management and production performance of Cobb-500 broiler parent stock under open housing system. Asian Australas. *J. Biosci. Biotechnol*. 1(1): 66-72.

Hasyim AR. 2015. Performa hasil persilangan ayam kampung ras pedaging dengan *Pelung* sentul pada umur 0-11 minggu (Thesis). [Bogor: (Indonesia)]: Institut Pertanian Bogor.

Henuk YL, Bakti D. 2018. Benefits of Promoting Native Chickens for Sustainable Rural Poultry Development in Indonesia. Mohammad Basyuni, S. Hut., M.Si., Ph.D., Prof. Dr. Ir. Elisa Julianti, M.Si, editors. Conference Proceeding of Seminar Ilmiah Nasional Dies Natalis USU-64. Sumatera Utara (Indones): University of Sumatera Utara. Pp. 69-76.

Iskandar S. 2017. Petunjuk tenis produksi ayam lokal pedaging unggul (Program Perbibitan Tahun 2017). Edisi 2017. Bogor (Indonesia): Pusat Penelitian dan Pengembangan Peternakan. Pp. 1-43.

Joller S, Bertschinger F, Kump E, Spiri A, von Rotz A, Schweizer-Gorgas D, Drogemuller C, Flury C. 2018. Crossed beaks in a local swiss chicken breed. *BMC Vet. Res*. 14: 68.

Kartika AA, Widayati KA, Burhanuddin, Ulfah A, Farajallah A. 2016. Eksplorasi preferensi masyarakat terhadap pemanfaatan ayam lokal di Kabupaten Bogor Jawa Barat. J. Ilmu Pertanian Indonesia (JIPI). 21(3): 180-185.

Kerje S, Sharma P, Gunnarsson U, Kim H, Bagchi S, Fredriksson R, Schutz K, Jensen P, von Heijne G, Okimoto R, Andersson L. 2004. The dominant200

Buletin Veteriner Udayana      Mahardhika *et al.*

white, dun and smoky color variants in chicken are associated with insertion/deletion polymorphisms in the *PMEL17* Gene. *Genetics*. 168: 1507-1518.

Mabelebele M, Norris D, Siwendu NA, Ng'ambi JW, Alabi OJ, Mbajiorgu CA. 2017. Bone morphometric parameters of the tibia and femur of indigenous and broiler chickens reared intensively. *Appl. Ecol. Environ. Res*. 15(4): 1387-1398.

Mariandayani HN, Darwati S, Sutanto E, Sinaga E. 2017. Peningkatan produktivitas ayam lokal melalui persilangan tiga rumpun ayam lokal pada generasi kedua. Prosiding Seminar Nasional Biologi 2017: Pendidikan Biologi untuk Masa Depan Bumi. Aceh (Indones): Jurusan Pendidikan Biologi, Universitas Syiah Kuala. Pp. 139-146.

Mariandayani HN, Solihin DD, Sulandari S, Sumantri C. 2013. Keragaman fenotipik dan pendugaan jarak genetik pada ayam lokal dan ayam broiler menggunakan analisis morfologi. *J. Vet*. 14(4): 475-484.

Nataamijaya AG. 2010. Pengembangan potensi ayam lokal untuk menunjang peningkatan kesejahteraan petani. *J. Litbang. Pertanian*. 29(4): 131 133.

Navara KJ, Anderson EM, Edwards ML. 2012. Comb size and color relate to sperm quality: a test of the phenotype-linked fertility hypothesis. *Behav. Ecol*. 12: 1036-1041

Ningsih R, Prabowo DW. 2017. Tingkat integrasi pasar ayam broiler di sentra produksi utama: studi kasus Jawa Timur dan Jawa Barat. *Bul. Ilmiah Litbang Perdagangan*. 11(2): 247-270.

Nurfadillah S, Rachmina D, Kusnadi N. 2018. Impact of trade liberization on Indonesian broiler competitiveness. *J. Indon. Trop. Anim. Agri*. 43(4): 429-437.

Nurhuda SA. 2017. Pertumbuhan generasi ketiga hasil persilangan ayam lokal dengan ayam ras pedaging sampai umur 12 minggu (Thesis). [Bogor: (Indones)]: Institut Pertanian Bogor.

Nuroso. 2010. Ayam kampung pedaging hari per hari. Jakarta, Penebar Swadaya.

Oldenbroek K, van der Waaij L. 2014. Textbook animal breeding: animal breeding and genetics for bsc students. Centre for Genetic Resources (Netherlands): The Netherlands and Animal Breeding and Genomics Centre.

Pauwels J, Coopman F, Cools A, Michiels J, Fremaut D, de Smet S, Janssens GPJ. 2015. Selection for growth performance in broiler chickens associates with less diet flexibility. *PLoS ONE*. 10(6): e0127819.

Paxton H, Tickle PG, Rankin JW, Codd JR, Hutchinson JR. 2014. Anatomical and biomechanical traits of broiler chickens across ontogeny. Part ii. Body segment inertial properties and muscle architecture of the pelvic limb. *Peer J*. 2: e473.

Pusat Data dan Sistem Informasi Pertanian. 2015. Outlook Komoditas Pertanian Sub Sektor Peternakan Daging Ayam. [accessed: April 29[th] 2019]. http://epublikasi.setjen.pertanian.go.id/download/file/213-outlook daging-ayam-2015

Rahman MR, Chowdhury SD, Hossain ME, Ahammed M. 2015. Growth and early laying performance of a broiler parent stock in an open-sided house under restricted feeding. *Bangladesh J. Anim. Sci*. 44(1): 40-45.

Semakula J, Lusembo P, Kugonza DR, Mutetikka D, Ssennyonjo J, Mwesigwa M. 2011. Estimation of live body weight using zoometrical measurements for improved marketing of indigenous chicken in the Lake Victoria basin of Uganda. *Livestock Res. Rural. Dev*. 23(8).

Shim MY, Karnuah AB, Mitchell AD, Anthony NB, Pesti GM, Anggrey SE. 2012. The effects of growth rate on leg morphology and tibia breaking strength, mineral density, mineral